\documentclass{cimento}

\usepackage{graphicx}
\usepackage{tikz}
\usetikzlibrary{calc}
\usetikzlibrary{patterns}
\usetikzlibrary{patterns}
\usetikzlibrary{positioning,decorations.pathmorphing,decorations.markings,snakes}
\usepackage{amsmath}
\usepackage{amsfonts}
\usepackage{amssymb}
\usepackage{physics}
\usepackage{slashed}
\usepackage{simplewick}
\usepackage{mathtools}
\usepackage{color}
\usepackage{siunitx}
\usepackage{feynmp-auto}

\title{Studying the production mechanisms of light meson resonances in two-pion photoproduction: a Regge approach}
\shorttitle{Production mechanisms of light meson resonances}
\author{\L.~Bibrzycki\from{ins:krakow1},
N.~Hammoud\from{ins:krakow2},
V.~Mathieu\from{ins:barcelona}\from{ins:madrid},
R.~J.~Perry\from{ins:barcelona}\thanks{perryrobertjames@gmail.com},
        \atque \\ 
A.~P.~Szczepaniak\from{ins:jlab}\from{ins:indiana1}\from{ins:indiana2}
}
\instlist{
\inst{ins:krakow1} Faculty of Physics and Applied Computer Science, AGH University of Krakow, al. A. Mickiewicza 30, 30-059 Kraków, Poland
\inst{ins:krakow2} Institute of Nuclear Physics, Polish Academy of Sciences, PL-31-342 Krak\'ow, Poland
\inst{ins:barcelona} Departament de F\'isica Qu\`antica i Astrof\'isica and Institut de Ci\`encies del Cosmos, Universitat de Barcelona,  E-08028 Barcelona, Spain
\inst{ins:madrid} Departamento de F\'isica Te\'orica, Universidad Complutense de Madrid and IPARCOS, E-28040 Madrid, Spain
\inst{ins:jlab} Theory Center, Thomas Jefferson National Accelerator Facility, Newport News, VA 23606, USA
\inst{ins:indiana1} Center for Exploration of Energy and Matter, Indiana University, Bloomington, IN 47403, USA
\inst{ins:indiana2} Department of Physics, Indiana University, Bloomington, IN 47405, USA
}

\begin{document}

\maketitle

\begin{abstract}
A calculation of the angular moments of two-pion photoproduction is presented.
The underlying theoretical model encodes the prominent $\rho(770)$ resonance and the expected leading background contribution coming from the Deck mechanism. The model contains a number of free parameters which are fit to experimental data. A good description of the angular moments is obtained.
\end{abstract}

\section{Introduction}
This proceedings focuses on a theoretical determination of  the angular moments of two (charged) pion photoproduction at low invariant mass of the $\pi^+\pi^-$ system. The angular moments are well-defined experimental observables which are bilinear in the partial waves of the two-pion subsystem. They are thus a key spectroscopic observable of interest. After presenting the theoretical model for two-pion photoproduction, the angular moments are computed, and compared with experimental results. 

\section{Kinematics and Preliminaries}
\label{sec:kinematics}
In this paper a theoretical model for the process
${\gamma}(q,\lambda_\gamma)+p(p_1,\lambda_1)\to \pi^+(k_1)+\pi^-(k_2)+p(p_2,\lambda_2)$ is described. The calculation is performed in the helicity frame. In this frame, the $\pi^+\pi^-$ system is at rest ($\mathbf{k}_1=-\mathbf{k}_2$) and the recoiling proton ($\mathbf{p}_2$) defines the negative $z$-axis. The $\pi^+$ recoils at an angle $\Omega^\text{H}=(\theta^\text{H},\phi^\text{H})$ to the $z$ axis. 
The scalar amplitudes in such a $2\to3$ process are maximally described by five independent kinematical variables. Thus in addition to the two angles for the $\pi^+$, the kinematic invariants 
$s=(p_1+q)^2$, $t=(p_1-p_2)^2$, and $s_{12}=(k_1+k_2)^2=m_{12}^2$ are used.
This set of kinematic variables, $(s,t,s_{12},\Omega^\text{H})$ is complete, in the sense that all other kinematic invariants may be computed by knowing these five. The notation for the phase-space and intensity is taken from Ref.~\cite{Mathieu:2019fts}. The differential cross section is defined as
\begin{equation}\label{eq:intentisty}
\frac{d\sigma}{dtdm_{12}d\Omega^\text{H}}=\kappa \sum_{\lambda_1\lambda_2 \lambda_\gamma}|\mathcal{M}_{\lambda_1\lambda_2\lambda_\gamma}(s,t,s_{12},\Omega^\text{H})|^2\,,
\end{equation}
where $\kappa$ contains the kinematical factors and  $\mathcal{M}_{\lambda_1\lambda_2\lambda\gamma}(s,t,s_{12},\Omega^\text{H})$ is the invariant matrix element. 
This paper will focus on a description of the unpolarized angular moments. These may be obtained from Eq.~\eqref{eq:intentisty} via
\begin{equation}
\expval{Y_{LM}}(s,t,s_{12})=\sqrt{4\pi}\int d\Omega^\text{H}\, \frac{d\sigma}{dtdm_{12}d\Omega^\text{H}}\text{Re}\,Y_{LM}(\Omega^\text{H})\,.
\end{equation}

\section{Description of Model}
\label{sec:model}
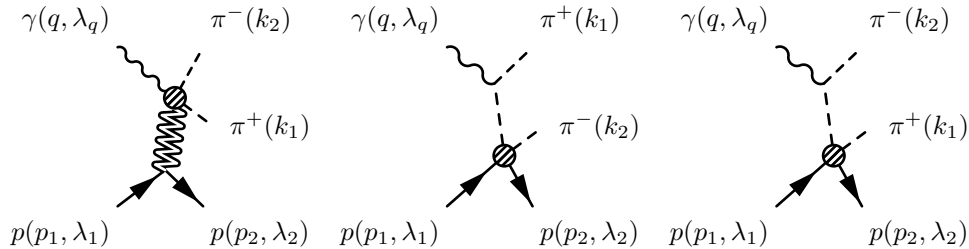
\begin{figure}
\centering
\vspace{20pt}
\begin{fmffile}{feynman_diagrams}
\begin{fmfgraph*}(40,60)
\fmfleft{l1,l2}
\fmfright{r1,r2,r3}
\fmf{fermion}{l1,v1,r1}
\fmf{photon}{l2,v2}
\fmf{dashes}{v2,r2}
\fmf{dbl_zigzag}{v1,v2}
\fmf{dashes}{v2,r3}
\fmfblob{0.2w}{v2}
\fmflabel{$p(p_1,\lambda_1)$}{l1}
\fmflabel{$\gamma(q,\lambda_q)$}{l2}
\fmflabel{$p(p_2,\lambda_2)$}{r1}
\fmflabel{$\pi^-(k_2)$}{r3}
\fmflabel{$\pi^+(k_1)$}{r2}
\end{fmfgraph*}
\hspace{80pt}
\begin{fmfgraph*}(40,60)
\fmfleft{l1,l2}
\fmfright{r1,r2,r3}
\fmf{fermion}{l1,v1,r1}
\fmf{photon}{l2,v2}
\fmf{dashes}{v1,r2}
\fmf{dashes}{v1,v2,r3}
\fmfblob{0.2w}{v1}
\fmflabel{$p(p_1,\lambda_1)$}{l1}
\fmflabel{$\gamma(q,\lambda_q)$}{l2}
\fmflabel{$p(p_2,\lambda_2)$}{r1}
\fmflabel{$\pi^-(k_2)$}{r2}
\fmflabel{$\pi^+(k_1)$}{r3}
\end{fmfgraph*}
\hspace{80pt}
\begin{fmfgraph*}(40,60)
\fmfleft{l1,l2}
\fmfright{r1,r2,r3}
\fmf{fermion}{l1,v1,r1}
\fmf{photon}{l2,v2}
\fmf{dashes}{v1,r2}
\fmf{dashes}{v1,v2,r3}
\fmfblob{0.2w}{v1}
\fmflabel{$p(p_1,\lambda_1)$}{l1}
\fmflabel{$\gamma(q,\lambda_q)$}{l2}
\fmflabel{$p(p_2,\lambda_2)$}{r1}
\fmflabel{$\pi^-(k_2)$}{r3}
\fmflabel{$\pi^+(k_1)$}{r2}
\end{fmfgraph*}
\end{fmffile}
\vspace{20pt}
\caption{Dominant contributions to two-pion photoproduction at small momentum transfer. With this approximation, it is possible to relate this $2\to3$ process to $2\to2$ processes.
}\label{fig:model}
\end{figure}

The global features of the angular moments, $\expval{Y_{LM}}$, measured in two-pion photoproduction can be understood as the result two competing mechanisms which may be separated based on their effect on the two-pion invariant mass distribution. In particular, the model incorporates both explicit low-energy two-pion resonances, and the leading continuum contribution from the ``Deck'' or ``Drell-S\"oding'' Mechanism~\cite{Drell:1961zz,Deck:1964hm,Soding:1965nh}. These contributions are represented diagramatically in Fig.~\ref{fig:model}. In the following subsections, these two mechanisms are described in more detail. 

\subsection{Resonant Model}
\label{sec:model:resonant}
The resonant model is constructed by assuming that two-pion resonances are produced via Reggeon-photon fusion, which subsequently decay into two pions. 
Symbolically, one may write
\begin{equation}
\mathcal{M}_{\lambda_1\lambda_2\lambda_\gamma}(s,t,s_{12},\Omega^\text{H})=\sum_{E} \Gamma_{\lambda_1\lambda_2}^{N\to EN}(t)R_E(s,t) i\mathcal{M}_{\lambda_\gamma}^{\gamma E\to \pi^+\pi^-}(s_{12},\Omega^\text{H})
\end{equation}
where $\Gamma_{\lambda_1\lambda_2}^{N\to EN}$ describes the nucleon-reggeon vertex, $R_E(s,t)$ is a Regge propagator and $i\mathcal{M}_{\lambda_\gamma}^{\gamma E\to \pi^+\pi^-}$ is the $\gamma E\to \pi^+\pi^-$ scattering amplitude. The Regge propagator is
\begin{equation}
R_E(s,t)=
\frac{\alpha^E(t)}{\alpha^E(0)}\frac{1+e^{-i\pi \alpha^E(t)}}{\sin\pi\alpha^E(t)}\bigg(\frac{s}{s_0}\bigg)^{\alpha^E(t)}
\end{equation}
where $\alpha_E(t)=\alpha_0^E+\alpha_1^E t$ is the Regge trajectory, and  $s_0=1~\si{GeV^2}$ is a mass scale introduced to fix the dimensions. In this study only $\mathbb{P}$, $f_2$ and $\rho/\omega$ exchanges are considered. The $t$-dependence of the vertices is not predicted from Regge theory and here is obtained from an effective Lagrangian approach. Residual $t$-dependence can be absorbed into the fitted parameters. Resonances are described by energy-dependent Breit-Wigner distributions following the formalism in Ref.~\cite{ParticleDataGroup:2022pth}.

The CLAS database~\cite{jlab_clas_database} contains experimental measurements of the two-pion angular moments from near threshold, $m_{12}\sim 0.4~\si{GeV}$, to $m_{12}\sim 1.4~\si{GeV}$. Clear evidence for the presence of meson resonances may be observed by studying the structures of the angular moments. In this proceedings, only resonances with a mass of less than $1~\si{GeV}$ will be considered. 
In this mass region, the PDG lists two $S$-wave  resonances, the $\sigma(500)$ and $f_0(980)$ and one $P$-wave resonance, the $\rho(770)$.

\subsection{Continuum Model}
\label{sec:model:continuum}
The Deck mechanism forms the continuum contribution to the model. For the reaction considered, it is given by 
\begin{equation}
\begin{split}
\mathcal{M}_{\lambda_1\lambda_2\lambda_\gamma}^\text{Deck,GI}&=\sqrt{4\pi\alpha}\bigg[\bigg(\frac{\epsilon(q,\lambda_\gamma)\cdot k_1}{q\cdot k_1}-\frac{\epsilon(q,\lambda_\gamma)\cdot(p_1+p_2)}{q\cdot (p_1+p_2)}\bigg)\beta(u_1){M}_{\lambda_1\lambda_2}^-(s_2,t;u_1)
\\
&-\bigg(\frac{\epsilon(q,\lambda_\gamma)\cdot k_2}{q\cdot k_2}-\frac{\epsilon(q,\lambda_\gamma)\cdot(p_1+p_2)}{q\cdot(p_1+p_2)}\bigg)\beta(u_2){M}_{\lambda_1\lambda_2}^+(s_1,t;u_2) \bigg]
\end{split}
\end{equation}
where 
$\beta(u_i)=\exp((u_i-u_i^\text{min})/\Lambda_\pi^2)$
are hadronic form factors introduced to suppress the Born term pion propagator at large $u_i=(q-k_i)^2$, $\Lambda_\pi=0.9~\si{GeV}$ and ${M}_{\lambda_1\lambda_2}^\pm$ is the scattering amplitude for the process $p+\pi^{*\pm }\to p+\pi^\pm$. Note that although this is a binary amplitude, the virtuality of the initial pion, $u_i$ implies that this amplitude is dependent on three kinematic invariants. In the limit that $u_i\to m_\pi^2$, these amplitudes may be related to elastic $\pi^\pm p$ scattering, for which there is a wealth of experimental and theoretical information. The form of the $\pi N$ amplitudes are taken from Ref.~\cite{Mathieu:2015gxa}. At low energy, this amplitude is described in terms of the SAID parameterization~\cite{Arndt:1994bu,Arndt:1995bj,Arndt:2003if} of $\pi N$ partial wave amplitudes, which are extracted from elastic $\pi N$ scattering data, while at high energies, the model interpolates to a Regge description of the data.

\section{Results and Discussion}
\label{sec:results_and_discussion}
\begin{figure}
    \centering
    \includegraphics[scale=0.25]{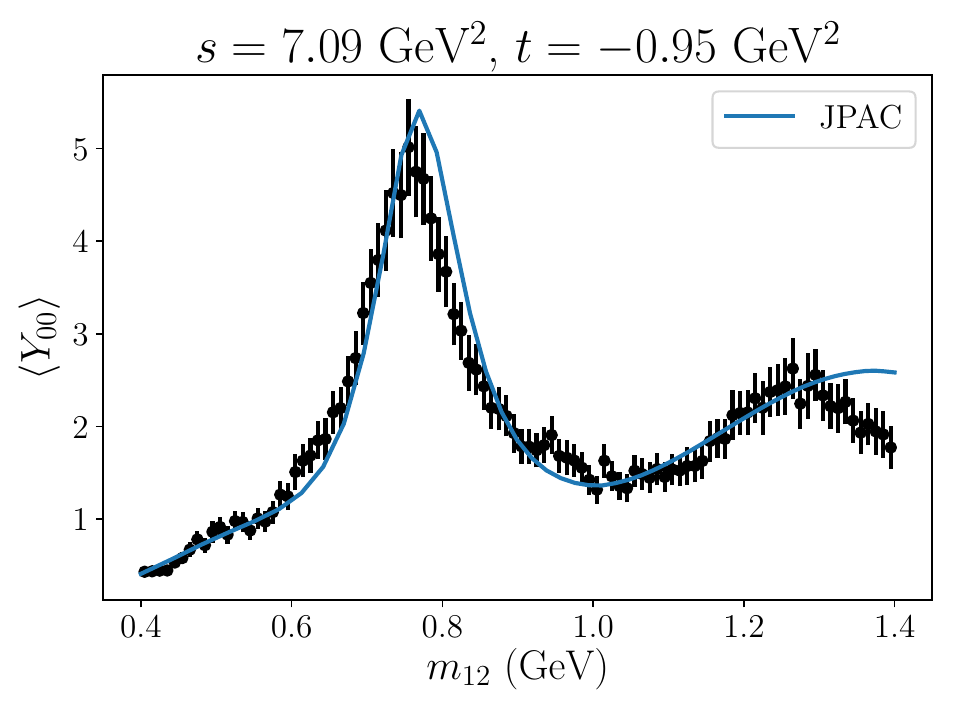}
    \includegraphics[scale=0.25]{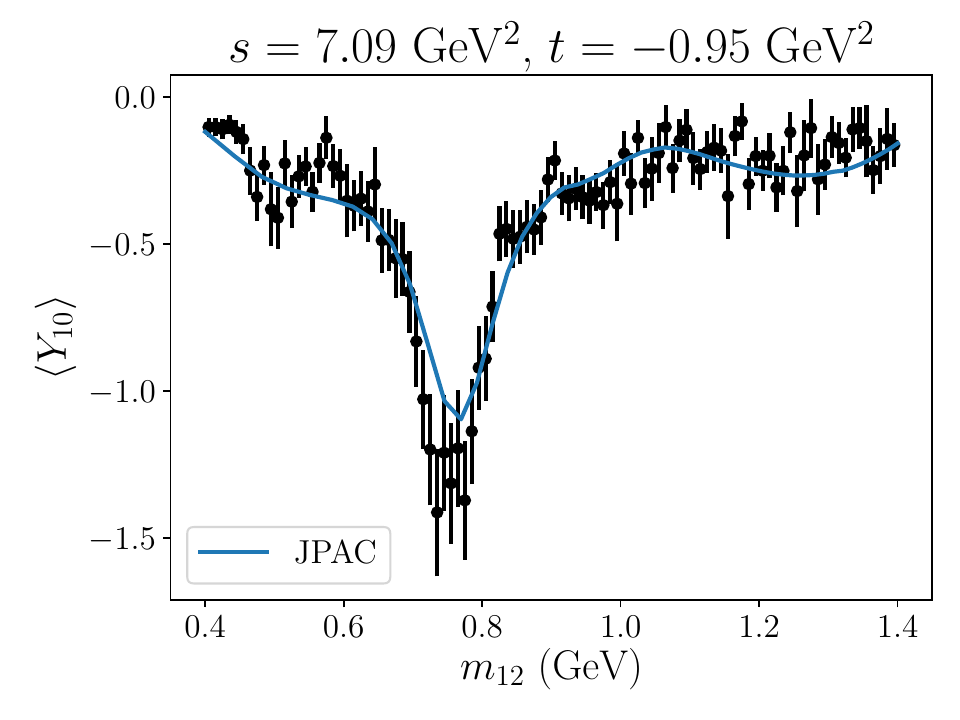}
    \includegraphics[scale=0.25]{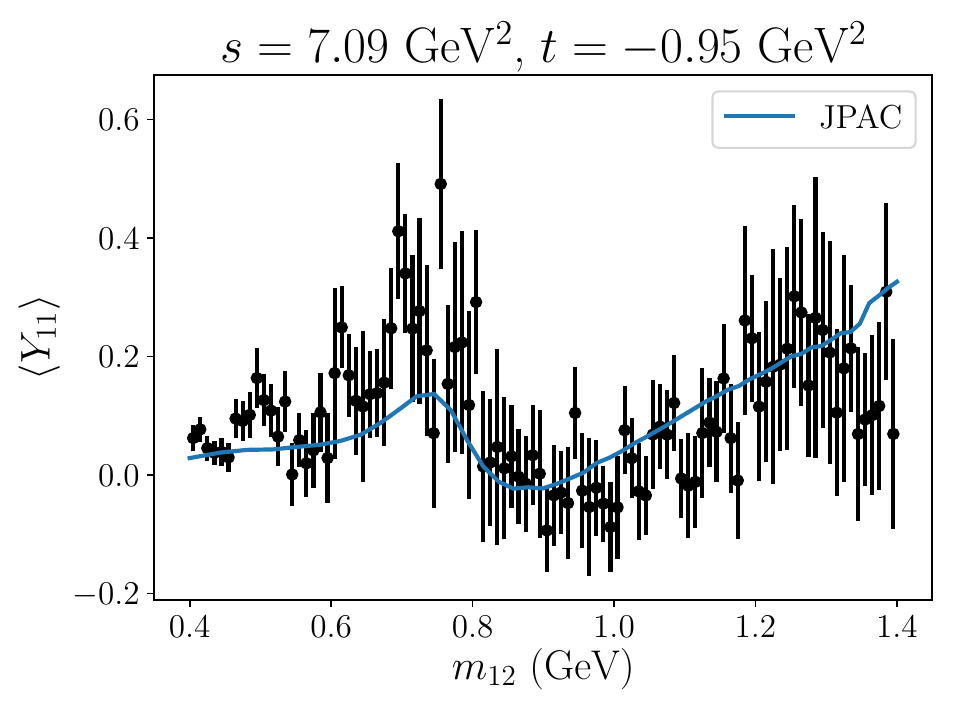}
    
    \includegraphics[scale=0.25]{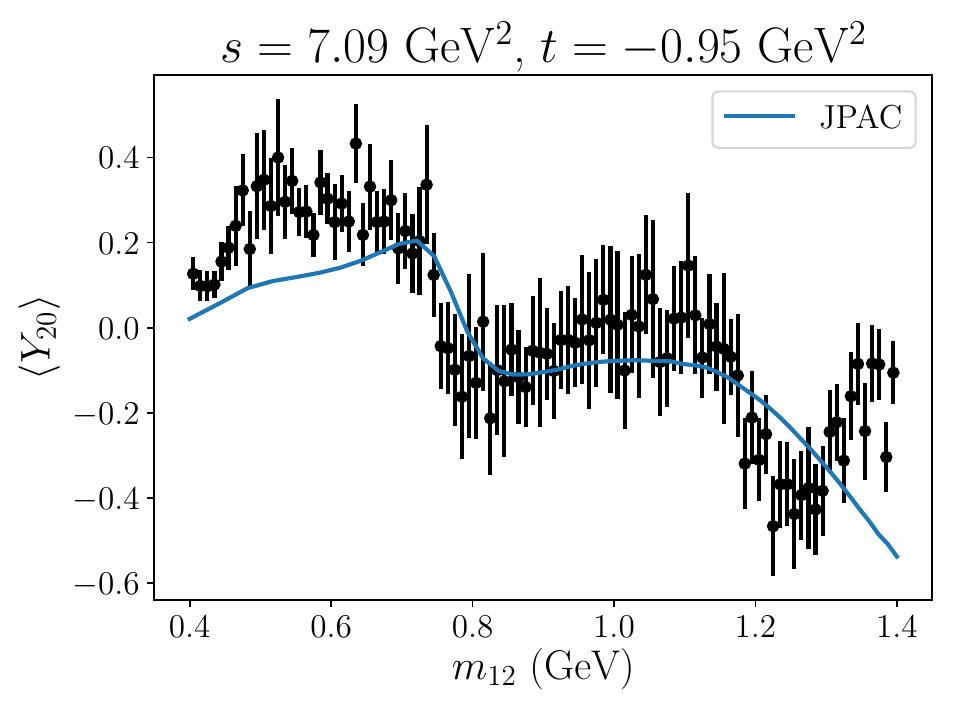}
    \includegraphics[scale=0.25]{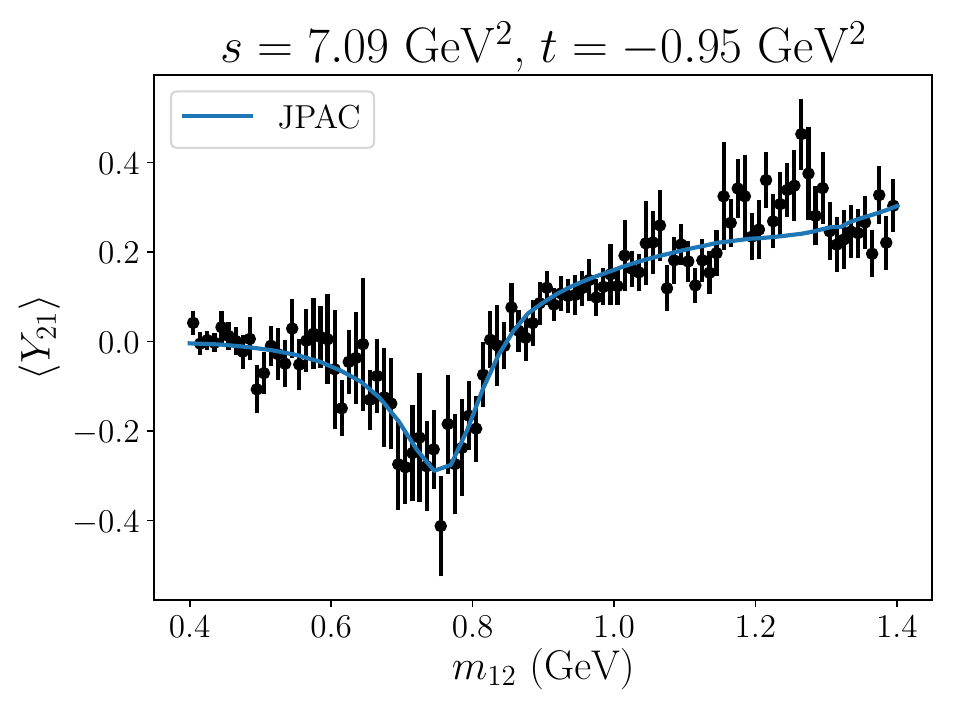}
    \includegraphics[scale=0.25]{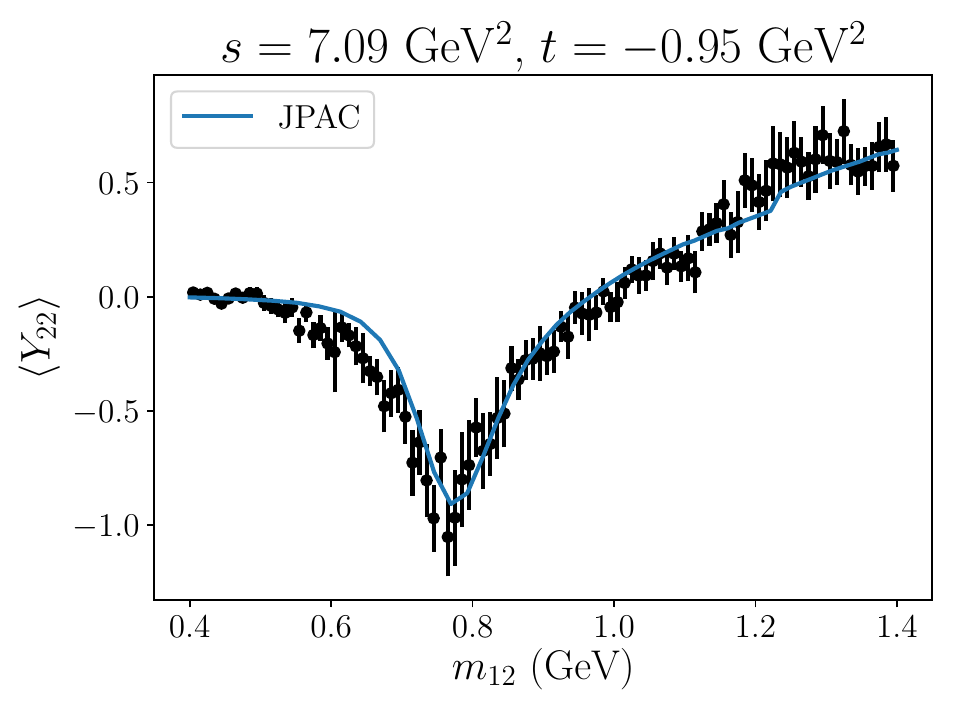}
    \caption{Comparing the preliminary predictions for angular moments of two-pion photoproduction with experimental data from Ref.~\cite{CLAS:2009ngd}. After fitting a number of free parameters, a good description of the experimental data is obtained.
    }
    \label{fig:angular_moments}
\end{figure}

Angular moments of two-pion photoproduction were computed using the model defined above and compared to the CLAS experimental data~\cite{CLAS:2009ngd}. In order to improve the agreement with experimental data, the relative weights of the partial waves were refit to the experimental data. The resulting fit to experimental data is shown in Fig.~\ref{fig:angular_moments}. A good fit to the angular moments is obtained, except possibly for  $m_{12}>1.2~\si{GeV}$. Note however, that in this region, there are several resonances which are expected to contribute, but which are not currently incorporated in the model. 

\begin{figure}
    \centering
    \includegraphics[scale=0.35]{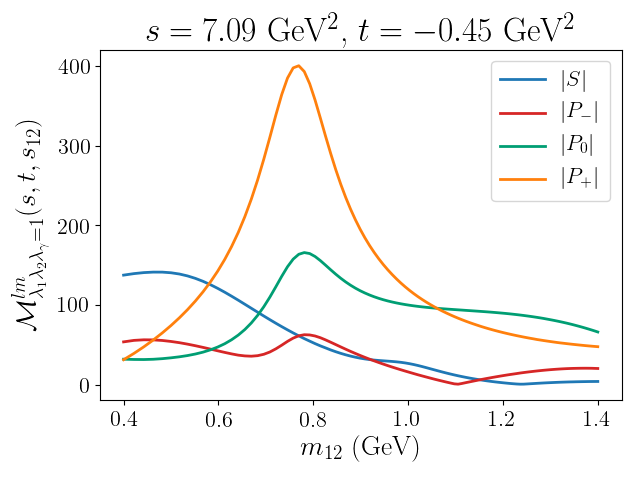}
    \includegraphics[scale=0.35]{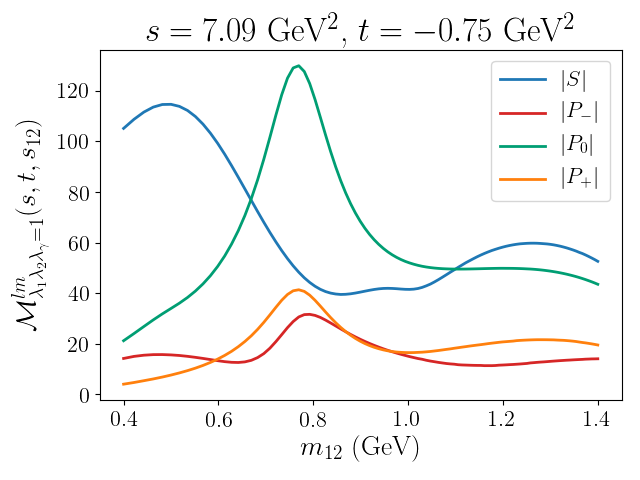}
    \caption{Preliminary prediction of the absolute value of $S$- and $P$-wave amplitudes. Note that at small $-t$, the hierachy of $P$-wave amplitudes is in accordance with SCHC, but at larger $-t$, the spin-flip amplitude $|P_0|$ appears to dominate. This behaviour can be understood from Regge theory.}
    \label{fig:partial_waves}
\end{figure}

The resulting absolute values of the partial-waves are shown in Fig.~\ref{fig:partial_waves}. The notation for the partial waves is taken from Ref.~\cite{CLAS:2009ngd}. In particular, $P_+$ denotes the $P$-wave partial wave amplitude with the same helicity as the incident photon, $P_0$ denotes the amplitude with one unit of helicity flip and $P_-$ denotes the amplitude with two units of helicity flip. It was observed in Ref.~\cite{Ballam:1970qn} that at small $-t$, the process $\gamma p \to\rho p$ proceeded primarily via the helicity conserving amplitude. This observation was called $s$-channel helicity conservation (SCHC). In the notation given here, this implies that $P_+$ dominates over $P_0$ and $P_-$. Regge theory implies that for small $-t/s$, the top and bottom vertices factorize, and the resulting $-t$-dependence of the top vertex is $\sim(-t)^{|\lambda_\gamma-\lambda_\rho|}$. Thus in the forward limit, only the helicity amplitude corresponding to $\lambda_\gamma=\lambda_\rho$ is expected to contribute. However, compared with Ref.~\cite{Ballam:1970qn}, the $-t$ measured at CLAS is larger. Due to the $-t$ dependence of the $P_0$ and $P_-$ amplitudes, one expects that at larger $-t$ SCHC should be increasingly violated, as these amplitudes are $-t$-enhanced. This behaviour is observed in the $t$-dependence  in Fig.~\ref{fig:partial_waves}. Note that at $-t=0.45~\si{GeV}$, the hierachy of partial waves is as one would predict from SCHC, with $|P_+|>|P_0|>|P_-|$. However, at larger $-t$, this hierachy is no-longer obeyed. 

\section{Conclusions}
This proceedings presented a theoretical model of two-pion photoproduction which incorporated the prominent $\rho(770)$ resonance and leading background contribution from the Deck mechanism. A good agreement between model and data was obtained after the relative weights of partial waves were fit to experimental data, suggesting that the model correctly identifies the production mechanisms and partial waves which are instrumental for the description of the angular moments.
This model can be applied to make predictions at photon energies relevant for the CLAS12 and GlueX experiments. 

\acknowledgments
This work contributes to the aims of the USDOE ExoHad Topical Collaboration, contract DE-SC0023598. In addition, the support of project PID2020-118758GB-I00, financed by the Spanish MCIN/ AEI/10.13039/501100011033/ is acknowledged. The support of project No. 2018/29/B/ST2/
02576 (National Science Center), partly financed by a Polish research project is also acknowledged. The authors thank their JPAC colleagues for useful conversations.

\bibliographystyle{varenna-bibtex/varenna}
\bibliography{varenna-bibtex/bibliography}

\end{document}